\documentclass[11pt]{article}
\usepackage{epsfig,amssymb}
\newcommand{\eq}[1]{(\ref{#1})}
\def\la{\langle }
\newcommand{\ra}{\rangle }
\def\l{\left(}
\def\r{\right)}

\begin{document}
\title{Constraints on ultra-high energy neutrinos from optically thick
astrophysical accelerators}

\author{D.S. Gorbunov$^a$, P.G. Tinyakov$^{a,b}$, and S.V. Troitsky$^a$ \\
\small\em
$^a$~Institute for Nuclear Research of the Russian Academy of
Sciences,\\
\small\em
60th October Anniversary Prospect 7a, 117312, Moscow, Russia;\\
\small\em
$^b$~Institute of Theoretical Physics, University of Lausanne,\\
\small\em CH-1015, Lausanne, Switzerland
}
\date{} 
\maketitle

\begin{abstract}
The $Z$-burst mechanism invoked to explain ultra-high energy cosmic rays
is severely constrained by measurements of the cosmic gamma-ray
background by EGRET. We discuss the case of optically thick sources
and show that jets and hot spots of active galaxies cannot provide the
optical depth required to suppress the photon flux. Other extragalactic
accelerators (AGN cores and sites of gamma ray bursts), if they are
optically thick, could be tested by future measurements of
the secondary neutrino flux.
\end{abstract}

\section{Introduction}
\label{sec:intro}
Recent observational data on ultra-high energy (UHE, $E\gtrsim
10^{19}$~eV) cosmic rays give a significant evidence for clustering in
their arrival directions \cite{clustering}. This fact suggests that
the observed extensive air showers are caused by particles created by
point-like sources.  Recently~\cite{BLLac} correlations of the arrival
directions of cosmic rays with BL Lac type objects -- certain active
galaxies located at cosmological distances -- were found (for earlier
discussion see Ref.~\cite{Farrar}). Taken seriously, these
data suggest that there exist particles which can travel for
cosmological distances unattenuated (without significant energy loss).

Among the Standard Model particles, only neutrinos can propagate
through the Universe unattenuated at ultra-high energy. However,
neutrino primaries are excluded by reconstruction of atmospheric
shower development \cite{FEnu,no-nu-primaries}. One of the ways out is to explore
the so-called ``$Z$-burst'' mechanism \cite{other-res} which works as
follows. The Hot Big Bang cosmology predicts the existence of cosmic
relic neutrino background.  Ultra-high energy neutrinos interact with
these background neutrinos very weakly, unless the energy is fine
tuned to the resonance \cite{Zburst} with $Z$ boson production in the
s-channel, that is,
\begin{equation}
E_{\rm res}\approx {4~{\rm eV}\over m_\nu}\cdot 10^{21}~{\rm eV},
\label{Eres}
\end{equation}
for the conventional cosmological model.\footnote{Note that
recently obtained limits on neutrino mass 
\cite{Fukugita} suggest $m_\nu\lesssim 1$~eV.} On resonance,
the interaction cross section increases significantly.  If the
resonant scattering takes place within $\sim 50$~Mpc from the Earth,
then secondary protons and photons produced in decays of virtual $Z$
bosons can serve as primaries of the extensive air showers.

In astrophysical accelerators, the neutrino production is usually
dominated by the two channels, namely, $p\gamma$ and $pp$ collisions,
where one proton has extremely high energy. If the collision energy in
the center-of-mass frame $E_{\rm cm}\gg1$~GeV, the total cross section
is saturated by multipion production and UHE neutrinos emerge mostly
as the products of charged pion decays. For $p\gamma$ processes the
collision energy in the center-of-mass frame may be smaller, $E_{\rm
cm}\sim1$~GeV, if UHE protons scatter off background soft photons
($E_\gamma\lesssim10^{-2}$~eV). In this case the cross section is
saturated by production of hadronic resonances (particular type of the
resonance depends on the energy in the center-of-mass frame). These
resonances decay into pions, protons and neutrons.  Then UHE neutrinos
appear as products of charged pion and neutron decays.
 
The generic feature of this mechanism is that it produces a certain
amount of protons and photons per each neutrino. This may lead to
contradiction with the observed fluxes if these particles leave the
source.  In the case of nucleons this statement is known as the
Waxman-Bahcall bound~\cite{Waxman} (see also Ref.~\cite{new-kalashev}): the charged cosmic ray (CR) flux
above $3\cdot 10^{18}$~eV is measured with relatively good accuracy
and implies that the sources have to be opaque for UHE nucleons.

A similar situation takes place for UHE photons which escape from the
source.  In the intergalactic space, the UHE photons give rise to
electromagnetic cascade transferring energy into less energetic photons which
propagate without attenuation \cite{transfer-to-EGRET}.  The
measurements of the flux of photons with $3\cdot
10^7~$eV$<E_\gamma<10^{11}$\ eV by EGRET \cite{egret} constrains UHE
neutrino flux in a similar way as the measurement of the charged CR
flux \cite{BerezinskyS}.  Though a detailed study of the propagation should take into
account a number of processes and numerical simulations are required
(see, e.g., Ref.~\cite{Semikoz}), even simple order-of-magnitude
estimates demonstrate that the most part of the primary photon energy flux
transfers to the EGRET energies.

A few ways were suggested which could help to overcome the EGRET
limits (recent discussions on these issues can be found, for instance,
in Refs.~\cite{N7,longR,Fodor}). 
One option is that the sources are more abundant at
distances closer to the Earth which, in average, results in less
energy losses \cite{N7}.  Alternatively, one may assume that sources are
optically thick for photons as well as nucleons, and only neutrinos
escape. In this paper we analyze the latter possibility.

\section{Optically thick part of the source}
\label{sec:damper}

Let us estimate the mass required to make the source optically thick
for photons. Denote the size of the absorption region (``damper'')
along the photon flux by $l_\|$ and the size in the transverse
direction by $l_\bot$. Let us consider first the energy loss by UHE
photons due to their interaction with protons. Optical thickness requires
that the size of the source is larger than the photon mean free
path. This condition can be written as
\begin{equation}
n_p>{1\over \sigma_{\gamma p}l_\|}\simeq
3\cdot10^5~\left({1~\!{\rm mb}\over\sigma_{\gamma p}}\right)
\left({1~\!{\rm kpc}\over l_\|}\right)~{\rm cm}^{-3},
\label{density}
\end{equation}
where $n_p$ is the proton number density in the damper and
$\sigma_{\gamma p}$ is the total cross section of UHE photons on
non-relativistic baryons. This cross section may be estimated by
extrapolating the low energy data~\cite{Groom:in} as
$\sigma_{\gamma p}\sim 1$~mb.  We are interested in the highest-energy
photons, $E_\gamma\gtrsim E_{\rm res}$.  Since $\sigma_{\gamma p}$
grows with energy, even larger proton density is required to suppress
less energetic photons, $E_\gamma<E_{res}$, if they are present in the
spectrum. 

\vspace{-0.5cm}
\phantom{.} From Eq.~(\ref{density}) 
one finds the total mass of the absorption 
region,
\begin{equation}
M_d\sim m_pn_pl_\bot^2l_\|>m_p{l_\bot^2\over\sigma_{\gamma p}}\simeq
10^{13}M_\odot \l {l_\bot\over 1~{\rm kpc}}\r^2 \left({1~{\rm
mb}\over\sigma_{\gamma p}}\right).
\label{total-mass}
\end{equation}

Inequalities (\ref{density}) and (\ref{total-mass}) actually
underestimate the proton density and the total mass of the absorption
region. The reason is that after each $\gamma p$ interaction,
approximately one third of the original photon energy flux is
transferred into neutral pions and, upon their subsequent decay, into
photons. Moreover, electrons and positrons from charged pion decays
carry 1/6 part of the original photon energy flux. Since strong
magnetic fields are usually present in astrophysical accelerators,
these electrons and positrons lose rapidly their energy into
synchrotron photons. All these secondary photons (both products of
$\pi^0$ decays and the synchrotron photons) affect the EGRET bound on
equal footing with the original photons. Thus, multiple scatterings
are required in order to further suppress the photon flux. The number
$N$ of these scatterings, by which the right hand sides of
Eqs.~\eq{density}, \eq{total-mass} should be multiplied, can be
estimated as follows.

Numerical analysis of the $Z$ burst scenario 
demonstrates~\cite{Yoshida,N7,longR} that
the required energy flux ${\cal{E}}\equiv E^2 j(E)$ in neutrinos at
energy $E\approx E_{\rm res}$ is certainly not less than
\begin{equation} 
{\cal{E}}_{\nu}(E_{\rm res})\simeq 5\cdot 10^4
~{\rm eV}~{\rm cm}^{-2}~{\rm s}^{-1}~{\rm sr}^{-1}\,.
\label{tot-nu-flux}
\end{equation}
If all these neutrinos are produced in $\gamma p$ or $pp$ collisions,
then they are accompanied by the photon flux
${\cal{E}}^0_\gamma(E_{\rm res}) \sim {\cal{E}}_{\nu}(E_{\rm res})$
(the exact ratio of photon and neutrino fluxes depends on the process
which dominates).  This flux has to be reduced by a factor of $\sim50$
in order not to overshoot the EGRET bound \cite{egret} of 
$\sim 10^3 ~{\rm eV}~{\rm cm}^{-2}~{\rm s}^{-1}~{\rm sr}^{-1}$. As at
each $\gamma p$ collision the photon energy flux, ${\cal{E}}_\gamma$,
becomes roughly two times smaller\footnote{In our estimate, we took
into account that UHE electrons and positrons lose their energy mostly
into synchrotron photons of high energy and neglected the fact that
$\sigma_{\gamma p}$ is smaller for these
photons in comparison with primary UHE photons. As a result, we
obtained a conservative estimate for $N$.}, the required number of
collisions is $N\gtrsim6$.

Let us turn now to possible sources of UHE particles.  The
active galaxies, and BL~Lac type objects in particular, provide
exceptional conditions for proton acceleration. The BL~Lacs are a
subclass of blazars without emission lines in spectra. The blazars are
active galaxies whose jets are pointing
towards us. It is believed that they differ from other active galaxies
only by viewing angle (see Ref.~\cite{blazars} for a review). So, some
information about structure and physics of these objects can be gained
from observation of usual active galaxies.

A typical active galaxy consists of three parts: the core, or active
galactic nucleus (AGN); the jets; and the lobes. It is believed that
the core is fuelled by a central black hole, which produces two
relativistic jets in opposite directions. The matter content of the
jets is not very well known, but it is often supposed that a
significant part of the jet energy is carried by protons. The jets
are highly collimated, with beaming angle of a few
degrees.\footnote{Since no correlation was found between the arrival
directions of UHECRs and positions of active galaxies seen by large
angles, we suppose that the neutrino flux is also highly
collimated, in the same direction as a jet.} The jets end in the
lobes, clouds of non-relativistic matter which are fuelled by the
particle fluxes of the jets. At the end of the jet the so-called ``hot
spot'' is formed. The typical size of a hot spot in the transverse
direction (and the typical width of a jet) is 
\cite{Meisenheimer} of order $1\div 10$~kpc. Among the active galaxies studied in
Ref.~\cite{Meisenheimer} only few hot spots exhibit $l_\bot\sim l_\|$,
while the most have $l_\|$ about 0.3~kpc.

We can now use the density and mass limits obtained above to rule out
blazar jets and hot spots as photon absorption sites. First, for
$l_\|\sim0.3\div3$~kpc, the required proton density~(\ref{density}) is
at least a few orders of magnitude higher than observed.\footnote{The
density of electrons in a hot spot of a particular active galaxy was
estimated \cite{edens} as $n_e\sim 500~$cm$^{-3}$. We assume that the
proton density is of the same order.} Second, for a
typical size $l_\bot$ of a hot spot (and jet) of order a few
kiloparsecs, the total mass of the {\em baryonic} part of the
absorption site should be of order $10^{14}M_\odot$ (see
Eq.~(\ref{total-mass})).  This exceeds by an order of magnitude the
mass of the heaviest known galaxies (extremely rare giant
galaxies)~\cite{GALAXIES} and by two orders of magnitude the mass of a
typical active galaxy~\cite{AGmasses}.

A protonic cloud is the most economic way to suppress the photon flux
because of large $\gamma p$ cross section and the fact that energy
transfers from electromagnetic channel into neutrinos. 
Alternatives to the protonic absorption region are {\it i}) a cloud
filled with a large number of soft photons and {\it ii}) a region with
very strong magnetic field.  The latter possibility is less attractive
because the energy remains in the electromagnetic channel, and for
less energetic secondary photons the interaction with the magnetic
field is strongly suppressed. So, let us consider an absorption site
filled by photons. In a way similar to
Eqs.~\eq{density},~\eq{total-mass}, we estimate the required photon
density,
\begin{equation}
n_\gamma>{1\over \sigma_{\gamma\gamma}l_\|}\simeq
5\cdot10^7~\left({6~\!\mu{\rm b}\over\sigma_{\gamma\gamma}}\right)
\left({1~\!{\rm kpc}\over l_\|}\right)~{\rm cm}^{-3},
\label{upper}
\end{equation}
and the total energy of the photons which fill the absorbing cloud:
\begin{equation}
E_d^\gamma\sim \omega_\gamma n_\gamma l_\bot^2l_\|>
\omega_\gamma{l_\bot^2\over\sigma_{\gamma\gamma}}
\simeq 2\cdot10^{13}M_\odot\l{\omega_\gamma\over 10~{\rm
MeV}}\r\l{l_\bot\over 1~{\rm kpc}}\r^2
\l{6~\mu{\rm b}\over\sigma_{\gamma\gamma}}\r\;.
\label{total-photon-energy}
\end{equation}
Here, $\omega_\gamma$ is the average energy of soft photons. For not
very energetic photons, such that the center-of-mass energy $E_{\rm
cm}\lesssim 10^{17}$~eV in photon-photon collisions, the dominant
process is double pair production with cross section of about
6~$\mu$b. For higher $E_{\rm cm}$ multipion production is important, 
but the corresponding cross section depends logarithmically
on the energy $\omega_\gamma$ and even at low energy, $E_{\rm cm}\sim
1$~TeV, the partial cross section for this channel is about ten
percent. Thus most part of the energy remains in the electromagnetic
channel: energy transfer into neutrinos in soft photonic cloud is
suppressed by multipion branching ratio, so the required number of
collisions $N$ is larger than in the case of protonic cloud.  Also,
$\sigma_{\gamma\gamma} \gtrsim 6~\mu$b and this cross section grows
with energy (at high $\omega_\gamma$) logarithmically, which means
that the total energy in photons, Eq.~\eq{total-photon-energy}, grows
with $\omega_\gamma$. The corresponding value of $E^\gamma_d$ exceeds
the mass of a galaxy, contrary to the astrophysical estimates
\cite{GALAXIES}. This means that the blazar jets and hot spots cannot
be optically thick due to photons with $\omega_\gamma\gtrsim1$~MeV.

In order to rule out the possibility to suppress UHE photon energy
flux by scattering off photons with energies $\omega_\gamma<1$~MeV we
estimate the soft photon density in the blazars. To this end, we
consider the most extensively studied blazar, Mrk~421.
The observational data~\cite{NED} concern
photons with frequencies $\nu>10^8$~Hz ($\omega_{\gamma}> 4\times
10^{-7}$~eV).  The flux of Mrk~421 as a function of $\nu$ exhibits a
plateau of $\sim 1$~Jy at $\nu=10^8\div10^{10}$~Hz, and decreases
monotonically at higher frequencies. Assuming that all photons
produced in the source escape without significant energy loss, and
that the total luminosity of a blazar is produced in an absorption
site of the size $\sim1$~kpc, we may estimate the photon number
density at $\nu=10^8\div10^{10}$~Hz,
\begin{equation}
n_\gamma\lesssim3\cdot10^3~{\rm cm}^{-3};
\label{soft-photons}
\end{equation}
at higher $\nu$ the number density is smaller. Eq.~\eq{soft-photons}
is consistent with the observational evidence for
$n_\gamma\simeq10^3~{\rm cm}^{-3}$ in a lobe of a particular active
galaxy \cite{CenB}. This density is much less than the one required
for optical thickness, Eq.~\eq{upper}, which means that the photons
with energies $\omega_\gamma < 1$~MeV cannot help to make AGN jets/hot
spots optically thick with respect to $\gamma\gamma$ interactions.

The only way to weaken the requirement~(\ref{upper}) is to fine-tune
the energy of the soft photons in order to get the maximal cross
section $\sigma\sim0.2$~b ($e^+e^-$ pair production at energy close to
threshold). In particular, UHE photons with energy of
$E_\gamma\sim10^{19}\div10^{23}$~eV scatter with such a large cross
section off soft photons with frequencies $\nu\sim10^3\div10^7$~Hz.
There are no experimental data on the number density of photons of
such low frequencies in blazars.  If at lower frequencies the number
density is the same as at $\nu\sim10^8\div10^{10}$~Hz,
Eq.~\eq{soft-photons}, photons with energies
$E_\gamma\sim10^{16}\div10^{23}$~eV can scatter off soft photons once
at most.  However, this cannot help to avoid EGRET bound because for
$e^+e^-$ production near threshold the energy remains in the
electromagnetic channel.

To summarize, the ultra-high energy neutrino flux required in the
Z-burst mechanism cannot be produced in blazar {\em jets} and {\em hot
spots}, since the latter do not have enough mass and density to
suppress the accompanying ultra-high energy photon flux. If not
suppressed by about two orders of magnitude, these photons would cascade
towards lower energies and overshoot the EGRET limit.

One can estimate parameters of a neutrino source required to suppress
the photon flux in the Z-burst scenario. As follows from
Eqs.~\eq{density}, \eq{total-mass}, \eq{upper},
\eq{total-photon-energy}, the linear size of the dense part of such a
source should not exceed 300~pc, and proton density $n_p$ has to be
larger than $10^6$~cm$^{-3}$, or photon density $n_\gamma$
significantly larger than
$10^9$~cm$^{-3}$ (we take into account the large number of 
collisions $N$). 
Among the known astrophysical objects capable to
accelerate particles up to ultra-high energies, only AGN cores,
neutron stars and gamma ray bursts could satisfy these criteria.

\section{Secondary neutrino flux}
\label{sec:nu}
Let us study now the features of the neutrino spectrum of a
hypothetical optically thick source of UHE neutrinos required for the
$Z$-burst mechanism.  Scatterings in a dense protonic
cloud (or in a cloud of very high energy photons) result in
a significant flux of lower-energy neutrinos which escape from the
source. We estimate the flux of these secondary neutrinos and confront
the ``$Z$-burst plus optically thick sources'' scenario with 
measured fluxes of high-energy neutrinos.

\begin{figure}
\centerline{\epsfig{file=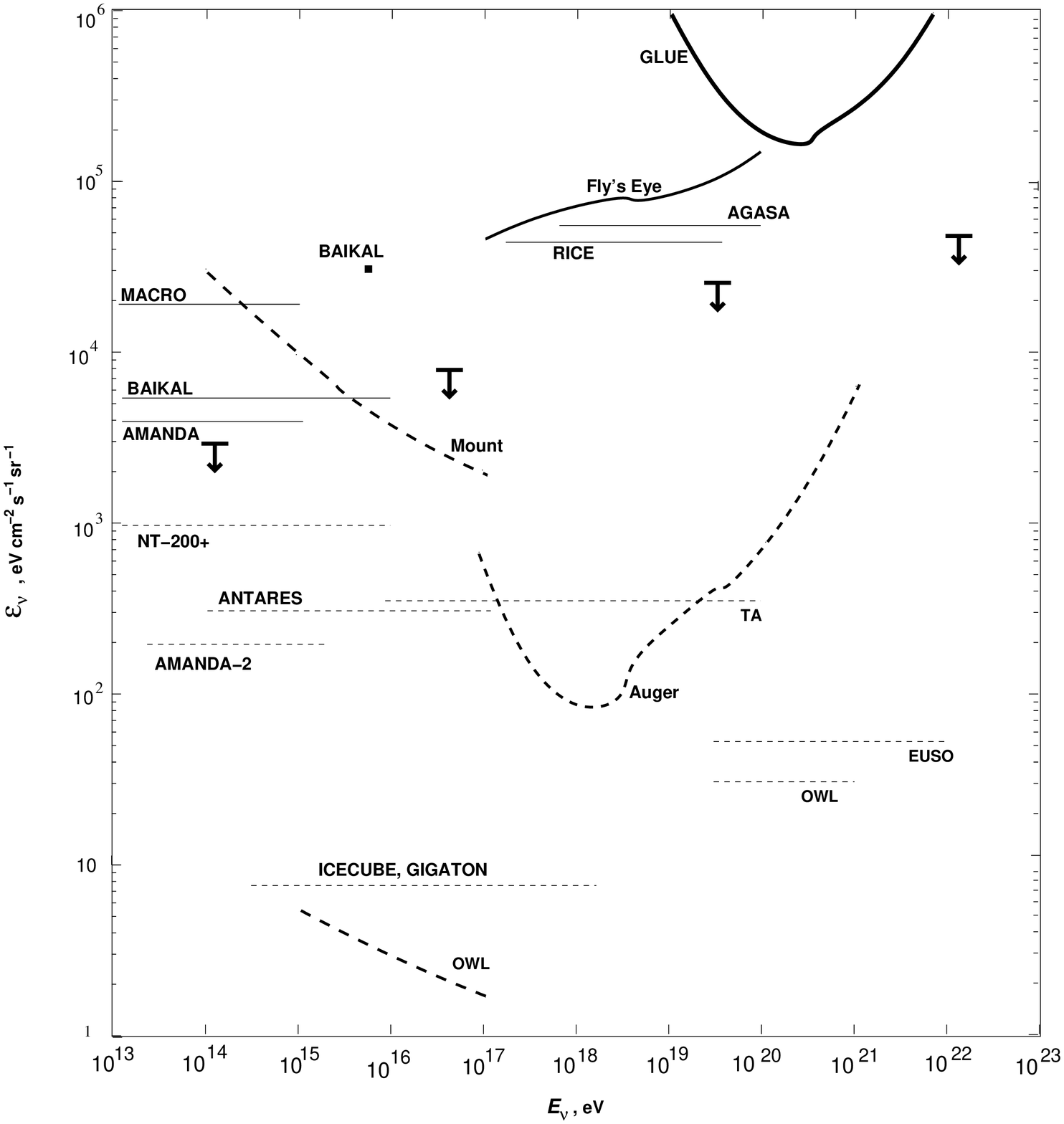, width=1.0\textwidth}}
\caption{Bold arrows represent the estimates of the total
$(\nu_e+\bar\nu_e+\nu_\mu+\bar\nu_\mu+\nu_\tau+\bar\nu_\tau)$ neutrino
flux from optically thick sources for \protect$E_{\rm res}=1.5\cdot
10^{22}$~eV (favoured by Ref.\protect\cite{Fodor}).  Solid lines
represent current bounds from MACRO \cite{MACRO}, Baikal
\cite{Baikal}, AMANDA \cite{AMANDA}, RICE \cite{RICE}, GLUE
\cite{GLUE}, AGASA \cite{AGASAnu}, and Fly's Eye \cite{FEnu}.
Projected sensitivities (dashed lines) of NT-200+ \cite{Baikal},
AMANDA-2 \cite{AMANDA-2}, ANTARES \cite{ANTARES}, IceCube
\cite{IceCube}, Gigaton \cite{Baikal}, Mount \cite{Mount}, Pierre
Auger \cite{Auger}, Telescope Array (TA) \cite{TA}, OWL \cite{OWL},
and EUSO \cite{EUSO} are also shown. For TA (10 stations), OWL, and
EUSO we assumed 10\% duty factor, one event per year detection
threshold, and neutrino-nucleon cross sections from
Ref.\cite{nu-n-cross-sec}. Thick lines (and a point) correspond to
``model-independent'' limits which assume monochromatic neutrino
fluxes, thin lines correspond to $E^{-2}$ assumed neutrino
spectrum. All limits were recalculated with $1:1:1$ flavour ratio
suggested by results on neutrino oscillations \cite{oscill}.}
\label{fig:nu}
\end{figure}

The required neutrino energy flux ${\cal{E}}_{\nu}(E_{\rm res})$
corresponds to a narrow bin, $E_{\rm res}-\Delta E \lesssim E \lesssim
E_{\rm res}+\Delta E$, where $\Delta E\approx0.03E_{\rm res}$ is the
width of the $Z$ resonance. To get a lower bound on the flux of the
secondary neutrinos we suppose that the original flux is zero at other
energies (which, of course, is not the case in a realistic source). We
assume that the relevant total $\gamma p$ and $\gamma\gamma$ cross
sections are saturated by multi-pion production. The average
multiplicity in high-energy collisions is a mild monotonically growing
function of the energy (see, e.g., Refs.~\cite{multiplicity}), and for
the energy scale of interest it can be approximated by a constant,
$\la n\ra\sim500$.  The average energy of each of the pions after the
first collision is $E_{\rm res}/\la n \ra$, and approximately one
third of the total energy is carried by neutral pions and the rest ---
by charged pions. Neutral pions decay to photons of lower energy
which, in turn, initiate new $p\gamma$ and/or $\gamma\gamma$
collisions (remember that $N\gtrsim 6$ collisions are required to
suppress the gamma ray flux below EGRET values).  Charged pions from
each collision decay to leptons, and the resulting neutrino flux is
easily estimated. We present the estimates of the high-energy
secondary neutrino flux in Fig.~\ref{fig:nu},
where the experimental bounds are also shown.  Note that UHE nucleons
which scatter off non-relativistic protons also produce secondary UHE
neutrinos. This results in additional contributions to the neutrino
flux which can be considered on equal footing with those from
$\gamma p$ and $\gamma\gamma$ collisions.  At very high energies, these
processes again are dominated by multipion production, and their
analysis is very similar to that of the $\gamma p$ case.

Though our estimates are fairly rough because exact details of
ultra-high energy and high-multiplicity processes are complicated and
sometimes unknown, our approach allows to check the viability of
various astrophysical sources to produce UHE neutrino flux required by
the ``$Z$-burst plus optically thick sources'' mechanism.  As is seen
from Fig.~\ref{fig:nu}, the projected sensitivities of 
future
experiments to neutrino flux at high energies 
are sufficient to check the possibility that the
enormous flux of photons required by the $Z$-burst mechanism is
suppressed in a protonic cloud, or in a cloud of very high energy 
photons, no matter what are the parameters
of the source. 

If the photonic cloud consists of less energetic photons, then
double pair production dominates, and details of the neutrino spectrum
depend on the magnetic field in the cloud. The energy flux of
neutrinos is suppressed then by multipion branching ratio with respect
to values shown in Fig.~1. However, the remaining secondary neutrino
flux is still within the projected sensitivity of high-energy neutrino
experiments.

On the other hand, if the absorption site is full of very soft photons
with $\nu\lesssim10^7$~Hz, the only contribution to the neutrino flux
would come from the first $\gamma p$ collision and only neutrinos at
energy $E_{\rm res}$ would be produced. Perspectives for detection of
these neutrinos were discussed in a number of papers on the $Z$-burst
mechanism (see, e.g., Refs.~\cite{Yoshida,Fodor,longR}).

\section{Conclusions}
\label{sec:concl}
Together with neutrinos required for the $Z$-burst explanation of UHE
cosmic rays, a certain number of energetic photons are produced in
astrophysical accelerators. Propagation of these photons in the
intergalactic space results in their reprocessing into softer gamma
rays with energies detectable by EGRET. If high-energy photons escape
from the sources, the resulting gamma-ray background is inconsistent
with EGRET observations for uniform source distribution. The sources, therefore, have to be opaque to
ultra-high energy photons. We estimated the required mass and the size
of the absorption region and found that jets and hot spots of active
galaxies (blazars, radio galaxies, etc.) cannot be optically thick to
UHE photons. They are, therefore, excluded as ``engines'' of the
Z-burst scenario. We then analyzed the flux of the secondary neutrinos
which are produced in the process of photon absorption in a
hypothetical optically thick source and obtained further
constraints. We argued that future measurements of neutrino flux at
high energies will
test the ``$Z$-burst plus optically thick protonic sources'' 
scenario. Our results are relevant not only for the $Z$-burst
mechanism but for any other mechanism which requires high flux of
photons in the source of cosmic rays.

\section*{Acknowledgements}
The authors are indebted to F.~Aharonian, O.~Catalano, G.~Domogatsky,
S.~Dubovsky, E.~Kozlovsky, M.~Libanov, K.~Postnov, A.~Ringwald,
V.~Rubakov, and D.~Semikoz for useful discussions and correspondence.
The work is supported in part by the program SCOPES of the Swiss NSF,
project No.~7SUPJ062239; by CPG and SSLSS grant 00-15-96626; by RFBR
grants 01-02-16710 (D.G.) and 02-02-17398 (D.G.\ and S.T.); by Swiss
NSF, grant 21-58947.99 (P.T.); by INTAS grants YSF 2001/2-142 (D.G.)
and YSF~2001/2-129 (S.T.).  This work has made use of the NASA/IPAC
extragalactic database \cite{NED}.

\end{document}